\documentclass[10pt,a4paper,twocolumn]{article}
\usepackage{graphicx}

\begin{document}
\title{A Method for Deriving the Dirac Equation from the Relativistic Newton's Second
Law}
\author{\small H. Y. Cui\footnote{E-mail: hycui@public.fhnet.cn.net}\\
\small Department of Applied Physics\\
\small Beijing University of Aeronautics and Astronautics\\
\small Beijing, 100083, China}
\date{\small \today}

\maketitle

\begin{abstract}
\small
The derivation becomes possible when we find a new formalism which connects
the relativistic mechanics with the quantum mechanics. In this paper, we
explore the quantum wave nature from the Newtonian mechanics by using a
concept: velocity field. At first, we rewrite the relativistic Newton's
second law as a field equation in terms of the velocity field, which
directly reveals a new relationship connecting to the quantum mechanics.
Next, we show that the Dirac equation can be derived from the field equation
in a rigorous and consistent manner.\\

\end{abstract}

In the past century, many attempts were made to address to understand the
quantum wave nature from the classical mechanics, however, much of the
connection with the classical physics is rather indirect. In this paper, we
propose a concept: velocity field, and show that the Dirac equation may be
derived from the relativistic Newton's second law in terms of the velocity
field in a rigorous manner.

According to the Newtonian mechanics, in a hydrogen atom, the single
electron revolves in an orbit about the nucleus, its motion can be described
with its position in an inertial Cartesian coordinate System $%
S:(x_1,x_2,x_3,x_4=ict)$. As the time elapses, the electron draws a spiral
path (or orbit), as shown in Fig.1(a) in imagination.

\begin{figure}[htb]
\includegraphics[bb=160 380 380 740,clip]{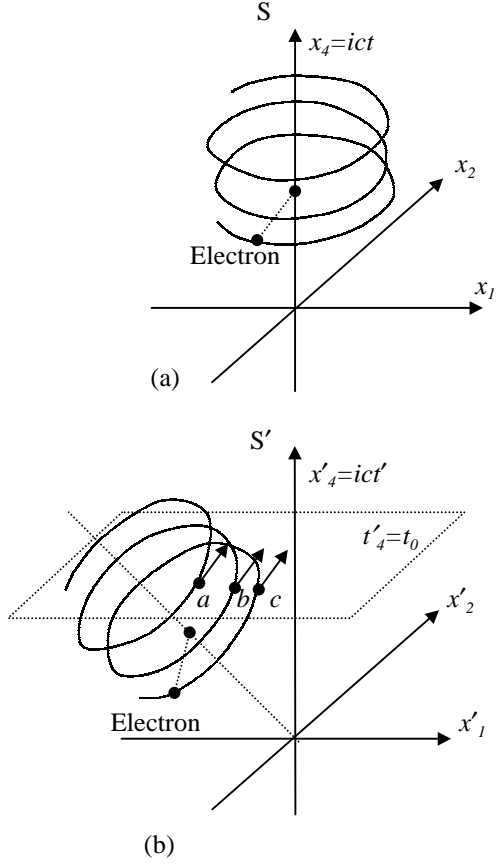}
\caption{The motion of the electron of hydrogen atom in 4-dimensional
space-time.}
\end{figure}

If the reference frame $S$ rotates through an angle about the $x_2$-axis in
Fig.1(a), becomes a new reference frame $S^{\prime }$, there will be a
Lorentz transformation linking the frames $S$ and $S^{\prime }$. Then in the
frame $S^{\prime }$, the spiral path of the electron tilts with respect to
the $x_4^{\prime }$-axis with the angle as shown in Fig.1(b). At one moment,
for example, $t_4^{\prime }=t_0$ moment, the spiral path pierces many points
at the plane $t_4^{\prime }=t_0$ , for example, the points labeled $a$, $b$
and $c$ in Fig.1(b), these points indicate that the electron can appear at
many points at the time $t_0$, in agreement with the concept of the
probability in quantum mechanics. This situation gives us a hint for
deriving quantum wave nature from the Newtonian mechanics.

Because the electron pierces the plane $t_4^{\prime }=t_0$ with 4-vector
velocity $u$, at every pierced point we can label a local 4-vector velocity
. The pierced points may be numerous if the path winds up itself into a cell
about the nucleus (due to a nonlinear effect in a sense), then the 4-vector
velocities at the pierced points form a 4-vector velocity field. It is noted
that the observation plane selected for the piercing can be taken at an
arbitrary orientation, so the 4-vector velocity field may be expressed in
general as $u(x_1,x_2,x_3,x_4=ict)$, i.e. the velocity $u$ is of a function
of position.

At every point in the reference frame $S^{\prime }$ the electron satisfies
the relativistic Newton's second law

\begin{equation}
m\frac{du_\mu }{d\tau }=qF_{\mu \nu }u_\nu  \label{1}
\end{equation}
the notations consist with the convention\cite{Harris}. Since the Cartesian
coordinate system is a frame of reference whose axes are orthogonal to one
another, there is no distinction between covariant and contravariant
components, only subscripts need be used. Here and below, summation over
twice repeated indices is implied in all case, Greek indices will take on
the values 1,2,3,4, and regarding the mass $m$ as a constant. As mentioned
above, the 4-vector velocity $u$ can be regarded as a 4-vector velocity
field, then

\begin{equation}
\frac{du_\mu }{d\tau }=\frac{\partial u_\mu }{\partial x_\nu }\frac{dx_\nu }{%
d\tau }=u_\nu \partial _\nu u_\mu  \label{2}
\end{equation}

\begin{equation}
qF_{\mu \nu }u_\nu =qu_\nu (\partial _\mu A_\nu -\partial _\nu A_\mu )
\label{3}
\end{equation}
Substituting them back into Eq.(\ref{1}), and re-arranging these terms, we
obtain

\begin{eqnarray}
u_\nu \partial _\nu (mu_\mu +qA_\mu ) &=&u_\nu \partial _\mu (qA_\nu ) 
\nonumber \\
&=&u_\nu \partial _\mu (mu_\nu +qA_\nu )-u_\nu \partial _\mu (mu_\nu ) 
\nonumber \\
&=&u_\nu \partial _\mu (mu_\nu +qA_\nu )-\frac 12\partial _\mu (mu_\nu u_\nu
)  \nonumber \\
&=&u_\nu \partial _\mu (mu_\nu +qA_\nu )-\frac 12\partial _\mu (-mc^2) 
\nonumber \\
&=&u_\nu \partial _\mu (mu_\nu +qA_\nu )  \label{4}
\end{eqnarray}
Using the notation

\begin{equation}
K_{\mu \nu }=\partial _\mu (mu_\nu +qA_\nu )-\partial _\nu (mu_\mu +qA_\mu )
\label{5}
\end{equation}
Eq.(\ref{4}) is given by

\begin{equation}
u_\nu K_{\mu \nu }=0  \label{6}
\end{equation}
Because $K_{\mu \nu }$ contains the variables $\partial _\mu u_\nu $, $%
\partial _\mu A_\nu $, $\partial _\nu u_\mu $ and $\partial _\nu A_\mu $
which are independent from $u_\nu $, then a solution satisfying Eq.(\ref{6})
is of

\begin{eqnarray}
K_{\mu \nu } &=&0  \label{7a} \\
\partial _\mu (mu_\nu +qA_\nu ) &=&\partial _\nu (mu_\mu +qA_\mu )
\label{7b}
\end{eqnarray}
The above equation allows us to introduce a potential function $\Phi $ in
mathematics, further set $\Phi =-i\hbar \ln \psi $, we obtain a very
important equation

\begin{equation}
(mu_\mu +qA_\mu )\psi =-i\hbar \partial _\mu \psi  \label{8}
\end{equation}
where $\psi $ representing the wave nature may
be a complex mathematical function, its physical meanings will be determined
from experiments after the introduction of the Planck's constant $\hbar $.

Multiplying the two sides of the following familiar equation by $\psi $

\begin{equation}
-m^2c^2=m^2u_\mu u_\mu  \label{9}
\end{equation}
which is valid at every point in the 4-vector velocity field, and using Eq.(%
\ref{8}), we obtain

\begin{eqnarray}
-m^2c^2\psi &=&mu_\mu (-i\hbar \partial _\mu -qA_\mu )\psi  \nonumber \\
&=&(-i\hbar \partial _\mu -qA_\mu )(mu_\mu \psi )-[-i\hbar \psi \partial
_\mu (mu_\mu )]  \nonumber \\
&=&(-i\hbar \partial _\mu -qA_\mu )(-i\hbar \partial _\mu -qA_\mu )\psi 
\nonumber \\
&&-[-i\hbar \psi \partial _\mu (mu_\mu )]  \label{10}
\end{eqnarray}
According to the continuity condition for the electron motion

\begin{equation}
\partial _\mu (mu_\mu )=0  \label{11}
\end{equation}
we have

\begin{equation}
-m^2c^2\psi =(-i\hbar \partial _\mu -qA_\mu )(-i\hbar \partial _\mu -qA_\mu
)\psi  \label{12}
\end{equation}
It is known as the Klein-Gordon equation.

On the condition of non-relativity, the Schrodinger equation can be derived
from the Klein-Gordon equation \cite{Schiff}(P.469).

However, we must admit that we are careless when we use the continuity
condition Eq.(\ref{11}), because, from Eq.(\ref{8}) we obtain

\begin{equation}
\partial _\mu (mu_\mu )=\partial _\mu (-i\hbar \partial _\mu \ln \psi
-qA_\mu )=-i\hbar \partial _\mu \partial _\mu \ln \psi  \label{13}
\end{equation}
where we have used the Lorentz gauge condition. Thus from Eq.(\ref{9}) to
Eq.(\ref{10}) we obtain

\begin{eqnarray}
-m^2c^2\psi &=&(-i\hbar \partial _\mu -qA_\mu )(-i\hbar \partial _\mu -qA_\mu
)\psi \nonumber \\
&&+\hbar ^2\psi \partial _\mu \partial _\mu \ln \psi \label{14}
\end{eqnarray}
This is of a complete wave equation for describing accurately the motion of
the electron. The Klein-Gordon equation is a linear equation so that the principle 
of superposition remains valid, however with the addition of the last term of Eq.(\ref{14}), 
Eq.(\ref{14}) turns to display chaos.

In the following we shall show the Dirac equation from Eq.(\ref{8}) and Eq.(%
\ref{9}). From Eq.(\ref{8}), the wave function can be given in integral form
by

\begin{equation}
\Phi =-i\hbar \ln \psi =\int\nolimits_{x_0}^x(mu_\mu +qA_\mu )dx_\mu +\theta
\label{15}
\end{equation}
where $\theta $ is an integral constant, $x_0$ and $x$ are the initial and
final points of the integral with an arbitrary integral path. Since the
Maxwell's equations are gauge invariant, Eq.(\ref{8}) should preserve
invariant form under a gauge transformation specified by

\begin{equation}
A_\mu ^{\prime }=A_\mu +\partial _\mu \chi ,\quad \psi ^{^{\prime
}}\leftarrow \psi  \label{16}
\end{equation}
where $\chi $ is an arbitrary function. Then Eq.(\ref{15}) under the gauge
transformation is given by

\begin{eqnarray}
\psi ^{^{\prime }} &=&\exp \left\{ \frac i\hbar \int\nolimits_{x_0}^x(mu_\mu
+qA_\mu )dx_\mu +\frac i\hbar \theta \right\} \exp \left\{ \frac i\hbar
q\chi \right\}  \nonumber \\
&=&\psi \exp \left\{ \frac i\hbar q\chi \right\}  \label{17}
\end{eqnarray}
The situation in which a wave function can be changed in a certain way
without leading to any observable effects is precisely what is entailed by a
symmetry or invariant principle in quantum mechanics. Here we emphasize that
the invariance of velocity field is hold for the gauge transformation.

Suppose there is a family of wave functions $\psi ^{(j)},j=1,2,3,...,N,$
which correspond to the same velocity field denoted by $P_\mu =mu_\mu $,
they are distinguishable from their different phase angles $\theta $ as in
Eq.(\ref{15}). Then Eq.(\ref{9}) can be given by

\begin{equation}
0=P_\mu P_\mu \psi ^{(j)}\psi ^{(j)}+m^2c^2\psi ^{(j)}\psi ^{(j)}  \label{18}
\end{equation}
Suppose there are matrices $a_\mu $ which satisfy

\begin{equation}
a_{\nu lj}a_{\mu jk}+a_{\mu lj}a_{\nu jk}=2\delta _{\mu \nu }\delta _{lk}
\label{19}
\end{equation}
then Eq.(\ref{18}) can be rewritten as

\begin{eqnarray} 
0 &=&a_{\mu kj}a_{\mu jk}P_\mu \psi ^{(k)}P_\mu \psi ^{(k)}  \nonumber \\
&&+(a_{\nu lj}a_{\mu jk}+a_{\mu lj}a_{\nu jk})P_\nu \psi ^{(l)}P_\mu \psi
^{(k)}|_{\nu \geq \mu ,when\nu =\mu ,l\neq k}  \nonumber \\
&&+mc\psi ^{(j)}mc\psi ^{(j)}  \nonumber \\
&=&[a_{\nu lj}P_\nu \psi ^{(l)}+i\delta _{lj}mc\psi ^{(l)}][a_{\mu jk}P_\mu
\psi ^{(k)}-i\delta _{jk}mc\psi ^{(k)}] \nonumber \\
\label{20}
\end{eqnarray}
Where $\delta _{jk}$ is the Kronecker delta function, $j,k,l=1,2,...,N$. For
the above equation there is a special solution given by

\begin{equation}
\lbrack a_{\mu jk}P_\mu -i\delta _{jk}mc]\psi ^{(k)}=0  \label{21}
\end{equation}

There are many solutions for $a_\mu $ which satisfy Eq.(\ref{19}), we select
a set of $a_\mu $ as

\begin{eqnarray}
N &=&4,\quad a_\mu =\gamma _\mu \quad (\mu =1,2,3,4)  \label{22a} \\
\gamma _n &=&-i\beta \alpha _n\quad (n=1,2,3),\quad \gamma _4=\beta
\label{22b}
\end{eqnarray}
where $\gamma _\mu ,\alpha $ and $\beta $ are the matrices defined in the
Dirac algebra\cite{Harris}(P.557). Substituting them into Eq.(\ref{21}), we
obtain

\begin{equation}
\lbrack ic(-i\hbar \partial _4-qA_4)+c\alpha _n(-i\hbar \partial
_n-qA_n)+\beta mc^2]\psi =0  \label{23}
\end{equation}
where $\psi $ is an one-column matrix about $\psi ^{(k)}$.

Let index $s$ denote velocity field, then $\psi _s(x)$ whose four component
functions correspond to the same velocity field $s$ may be regarded as the
eigenfunction of the velocity field $s$, it may be different from the
eigenfunction of energy. Because the velocity field is an observable in a
physical system, in quantum mechanics we know that $\psi _s(x)$
constitute a complete basis in which arbitrary function $\phi (x)$ can be
expanded in terms of them

\begin{equation}
\phi (x)=\int C(s)\psi _s(x)ds  \label{24}
\end{equation}
Obviously, $\phi (x)$ satisfies Eq.(\ref{23}). Eq.(\ref{23}) is well known
as the Dirac equation.

Alternatively, another method to show the Dirac equation is more
traditional: At first, we show the Dirac equation of free particle by
employing plane waves, we easily obtain Eq.(\ref{23}) on the condition of $%
A_\mu =0$; Next, adding electromagnetic field, plane waves are valid in any
finite small volume with the momentum of Eq.(\ref{8}) when we regard the
field to be uniform in the volume, so the Dirac equation Eq.(\ref{23}) is
valid in the volume even if $A_\mu \neq 0$, plane waves constitute a
complete basis in the volume; Third, the finite small volume can be chosen
to locate at anywhere, then anywhere have the same complete basis, therefore
the Dirac equation Eq.(\ref{23}) is valid at anywhere.

Of course, on the condition of non-relativity, the Schrodinger equation can
be derived from the Dirac equation \cite{Schiff}(P.479).

By further calculation, The Dirac equation can arrive at the Klein-Gordon equation 
with an additional term which represents the effect of spin, this term is just the last term 
in Eq.(\ref{13}) in a sense.

But, do not forget that the Dirac equation is a special solution of Eq.(\ref
{20}), therefore we believe there are some quantum effects beyond the Dirac
equation. 

We do not know exactly what kind of the path of the electron in a hydrogen
atom is, so the illustration of Fig.1 is an imaginary one for visualizing
the motion of the electron. But we know that the electron path will pierce
many points at any observation time plane like $t_4^{\prime }=t_0$ for
arbitrary reference frame $S^{\prime }$ if the path or orbit exists in the
4-dimensional space-time, the points may be numerous. Therefore there is a
4-vector velocity field for the motion of the electron. The 4-vector
velocity field is a key concept for our deduction.

It follows from Eq.(\ref{8}) that the path of particle is analogous to
''lines of electric force'' in 4-dimensional space-time. In the case that
the Klein-Gordon equation is valid, i.e. Eq.(\ref{11}) is valid, at any
point, the path can have but one direction (i.e. the local 4-vector velocity
direction), hence only one path can pass through each point of the
space-time. In other words, the path never intersects itself when it winds
up itself into a cell about the nucleus. No path originates or terminates in
the space-time. But, in general, the divergence of the 4-vector velocity
field does not equal to zero, as indicated in Eq.(\ref{13}).

In the Minkowski's space, every particle has its 4-vector velocity with the same magnitude $|u|=ic$. We now consider a situation where the electron revolves about the nucleus in
the 2D plane $x_1-x_4$ as shown in Fig.2. The first question: how the electron
surpasses the nucleus when both the electron and nucleus have their own
4-vector velocities with the same magnitude $|u|=ic$ ? According to the
theorem of relativistic addition of velocities, when the electron revolves
the nucleus at the light speed with respect to the nucleus, no matter what
speed of the nucleus is, the electron speed seem still being at the light
speed in laboratory reference system. The second question: the electron will
intersects itself path in the plane, for example, point P in Fig.2, how to
explain ? Certainly, because of the intersection, the electron collides with
itself, there is the divergence of 4-vector velocity field at every point,
as indicated by Eq.(\ref{13}). This new character profoundly accounts for
the quantum wave natures such as spin effect\cite{Cui0173}.

\begin{figure}[htb]
\includegraphics[bb=175 550 350 735,clip]{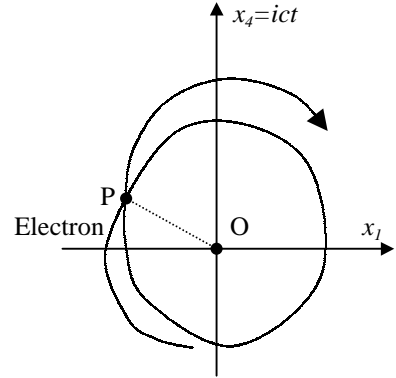}
\caption{The electron revolves about the nucleus in the plane $x_1-x_4$}
\end{figure}

In conclusion, the path of the electron of hydrogen atom should wind up
itself into a cell about the nucleus in 4-dimensional space-time, therefore
there is a 4-vector velocity field for describing the motion of the
electron. In terms of the 4-vector velocity field, the relativistic Newton's
second law can be rewritten as a wave field equation. By this, the
Klein-Gordon equation, Schrodinger equation and Dirac equation can be
derived from the relativistic Newtonian mechanics on different conditions,
respectively.

\end{document}